\documentclass[doublecol]{epl2} 
\usepackage[latin1]{inputenc}
\usepackage{graphicx}
\usepackage{amsmath}
\usepackage{latexsym}
\usepackage{amsbsy}
\usepackage{color}  

\title{Observation of the 2p$\boldsymbol{_{3/2}\to}$2s$\boldsymbol{_{1/2}}$ intra-shell transition in He-like uranium}
\shorttitle{Observation of the 2p$\to$2s intra-shell transition in He-like U} 

\author{M. Trassinelli\inst{1,2}\thanks{E-mail: \ martino.trassinelli@insp.jussieu.fr} \and A.~Kumar\inst{2,3} \and H.F.~Beyer\inst{2} \and P.~Indelicato\inst{4} \and R.~Märtin\inst{2,5} \and R.~Reuschl\inst{1,2} \and  Y.S.~Kozhedub\inst{6} \and C.~Brandau\inst{2} \and H.~Bräuning\inst{2} \and S.~Geyer\inst{2} \and A.~Gumberidze\inst{2} \and S.~Hess\inst{2} \and P.~Jagodzinski\inst{7} \and C.~Kozhuharov\inst{2} \and D.~Liesen\inst{2} \and U.~Spillmann\inst{2} \and S.~Trotsenko\inst{2} \and G.~Weber\inst{2,5} \and D.F.A.~Winters\inst{2,5} \and Th.~Stöhlker\inst{2,5}
 }
\shortauthor{M. Trassinelli \etal}

\institute{                    
  \inst{1} CNRS, UMR 7588, INSP, Campus Boucicaut, 140 rue de Lourmel, Paris, 75015 France, Université Pierre et Marie Curie-UPMC, UMR 7588, INSP, Campus Boucicaut, 140 rue de Lourmel, Paris, 75015 France  \\
  \inst{2} GSI Helmholtzzentrum für Schwerionenforschung GmbH, Planckstraße 1,  64291 Darmstadt, Germany\\
  \inst{3} Nuclear Physics Division, Bhabha Atomic Research Centre, Mumbai - 400 085, India \\
  \inst{4} Laboratoire Kastler Brossel, {É}cole Normale Sup{é}rieure; CNRS; Universit{é} Pierre et Marie Curie-Paris 6 - Case 74, 4 Place Jussieu, 75005 Paris, France \\
  \inst{5} Physikalisches Institut, Universität Heidelberg - Philosophenweg 12, 69120 Heidelberg, Germany \\
  \inst{6}Department of Physics, St. Petersburg State University - Oulianovskaya 1, Petrodvorets, St. Petersburg 198504, Russia \\
  \inst{7} Institute of Physics, Jan Kochanowski University - ul. \'Swiketokrzyska 15, 25406 Kielce, Poland 

}

\pacs{32.30.Rj}{X-ray spectra}
\pacs{31.30.J-}{Relativistic and quantum electrodynamic effects in atoms, molecules, and ions}
\pacs{12.20.Fv}{Experimental tests}
\pacs{07.85.Nc}{X-ray and $\gamma$-ray spectrometers}
\pacs{32.10.Fm}{Fine and hyperfine structure}

\abstract{We present the first observation of the $1s2p\, ^3\!P_2 \to 1s2s\, ^3\!S_1$ transition in He-like uranium.
The experiment was performed at the internal gas-jet target of the ESR storage ring at GSI exploiting a Bragg crystal spectrometer  and a germanium solid state detector.
Using the $1s^22p\, ^2\!P_{3/2} \to 1s^22s\, ^2\!S_{1/2}$ transition in Li-like uranium as reference and the deceleration capabilities of the ESR storage rings, we obtained the first evaluation of the He-like heavy ion intra-shell transition energy.
}

\begin{document}

\maketitle

\begin{figure}
\begin{center}
\includegraphics[width=\columnwidth]{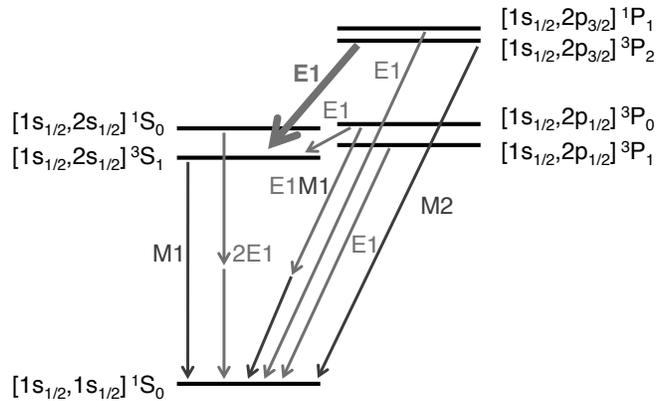}
\caption{Level scheme of helium-like uranium and major decay branches.
The bold arrow indicates the transition studied.}
\label{fig:scheme}
\end{center}
\end{figure}

He-like ions are the simplest multi-body atomic systems.
Investigations of these ions along their isoelectronic sequence up
to the heaviest species uniquely probe our understanding of
correlation, relativistic, and Quantum Electrodynamic (QED) effects.
In recent years, substantial progress in the investigations of these
fundamental systems has been achieved in the high-$Z$ region, 
in which the nucleus Coulomb field strength and electron velocity, 
both proportional to $Z \alpha$, are very high.
In
theory, benchmark calculations have been reported where even second
order QED effects were considered in a rigorous way for both the
ground-state as well as for the first excited states
\cite{Persson1996,Plante1994,Artemyev2005}. Experimentally, 
progress has been mainly achieved by a technique recently introduced
\cite{Marrs1995}, which allows 
for the isolation of the two-electron
contributions to the ionization potential of the ground-state
\cite{Marrs1995,Mokler2008,Gumberidze2004}. Here, the achieved
uncertainty already approaches the expected size of higher-order
QED contributions \cite{Mokler2008,Gumberidze2004}.
However, for the excited levels at high-$Z$ ($Z>54$) virtually no
experimental data on binding or transition energies are available.
Besides the general importance of such data for atomic structure
investigations, the great interest in more detailed information
about the excited levels in high-$Z$ He-like ions is motivated by
their relevance for the study of the influence of the electroweak
interaction on the atomic structure. Since many years, high-$Z$
He-like ions are under the discussion as an ideal candidates for the study of atomic
parity violation effects \cite{Schafer1989,Labzowsky2001}. 
The ability to observe parity-violating transitions critically depends on the relative binding energies of the excited states.
A level
crossing of two states with different parity, namely the $1s2p\,
^3\!P_0 $ and the $1s2s\, ^1\!S_0 $ levels, is predicted to occur
close to $Z=66$ and $Z=92$. However, a benchmark test of the atomic
structure theory for the excited states in He-like ions is still
pending. 
Therefore, a precise measurement of the $\Delta n=0$ $1s2p\, ^3\!P_2 \to 1s2s\, ^3\!S_1$
intra-shell transition energy is of particular importance.
(a partial level scheme
of a high-$Z$ He-like ion is shown in Fig.~\ref{fig:scheme}).
An attempt of an energy measurement
was reported for the $1s2p \to 1s2s$ intra-shell transition for
He-like uranium using the electron-beam ion trap (EBIT) at Lawrence
Livermore National Laboratory (LLNL) \cite{Beiersdorfer1996}. There,
the desired $1s2p\, ^3\!P_2 \to 1s2s\, ^3\!S_1$ transition could not
be identified unambiguously. This is in contrast to lighter ions up
to $Z=54$, where  energy measurements of the $1s2p \to 1s2s$
transitions have been performed by spectroscopy in the visible to
the far UV region
 \cite{Kukla1995,Artemyev2005,Martin1989}. For higher values of $Z$, information on excited states has been
obtained only via lifetime measurements of the $1s2p\, ^3\!P_0$ level
\cite{Indelicato1992,Toleikis2004,Munger1986}.

In this Letter we present the first clear identification of the $1s2p\, ^3\!P_2 \to 1s2s\, ^3\!S_1$
intra-shell transition in He-like uranium and its energy measurement. 
For the purpose of the experiment, we used a standard Ge(i) solid-state detector and a new Bragg spectrometer
specially designed for accurate spectroscopy of fast ions.
The two instruments are complementary: The Ge(i) detector has a high detection efficiency and
covers a wide spectral range with a moderate spectral resolving power. The focussing crystal spectrometer
serves as an accurate wavelength comparator in a narrow wavelength interval.
In concert, the present measurements allow for an unambiguous identification of the spectral lines
observed plus an accurate determination of the transition wavelength or energy.

The experiment was performed at the ESR storage ring at the GSI in
Darmstadt. 
Here, a beam of $\sim 4 × 10^7$ hydrogen-like uranium
ions was stored, cooled and decelerated to an energy of
43.57~MeV/u. The ion beam's momentum spread was close to $\Delta p
/ p \approx 10^{-5}$, and its width was about 2~mm. 
He-like excited ions
were formed by electron capture during the interaction of the ion
beam with a supersonic nitrogen gas-jet. The gas-jet had a width of
about 5~mm and a typical areal density of $10^{12}$ particles/cm$^2$,
which guaranteed single-collision conditions in the ion--target
interaction. At the selected velocity, electrons are primarily
captured into the shells with principal numbers $n \leq$ 20
\cite{Fritzsche2000,Ma2001,Eichler2007}.
This  allows for an efficient
population of the $1s2p\, ^3\!P_2$ level via cascade feeding.

The $1s2p\, ^3\!P_2$
excited state mainly decays in two ways (Fig.\ \ref{fig:scheme}): to the ground state via a
magnetic quadrupole (M2)  transition, with a branching ratio of
70\% and a decay time of $4.9 × 10^{-15}$~sec, and to the $1s2s\, ^3\!S_1$ state by an electric dipole (E1)
intra-shell transition, with a branching ratio of 30\% and a decay time of $1.2 × 10^{-14}$~sec
\cite{Indelicato2005a}. 
In the reference frame of the projectile,
the x-ray
photons arising from the $2\,  ^3\!P_2 \to 2\, ^3\!S_1$ transition
have  an energy of 4510~eV. 

The Ge(i) solid state detector and the Bragg crystal 
spectrometer were mounted under observation angles of $35^\circ$ and $90^\circ$, respectively. 
Both instruments were
separated from the ultra-high vacuum of the gas target chamber by
100$\mu$m-thick beryllium windows transparent for the low-energy x
rays.

\begin{figure}
\includegraphics[width=\columnwidth]{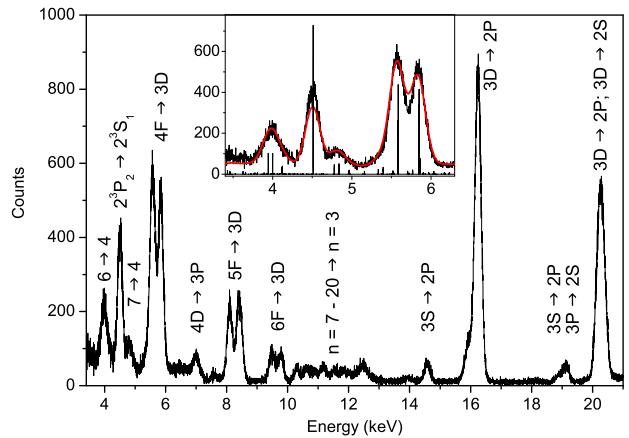}
\caption{The x-ray spectrum (color online), originating from 43.57 MeV/u He-like
uranium ions, as recorded by the Ge(i) detector. The energies correspond to the
emitter frame. The inset shows a zoom around the $^3\!P_2 \to ^3\!S_1$ transition with the vertical
bars and  the continuous (red) curve representing the results of a  full cascade calculation \cite{Stohlker1999}.}
\label{fig:ge_spectra}
\end{figure}

The Ge(i) detector crystal has a diameter of 16 mm and therefore
provides a relatively  large solid angle. An x-ray collimator was
mounted in front of the Ge(i) in order to limit the horizontal
acceptance angle thereby reducing the Doppler broadening to match the corresponding Doppler width with the intrinsic line width of the detector, amounting to 250 eV  at 5.9 keV photon energy.

A survey spectrum recorded by the Ge(i) detector is displayed in Fig.~\ref{fig:ge_spectra} with
prominent lines originating from $n' \to n$ transitions, with $n =
2-4$. 
The existence of lines originating from high $n'$ to $n=2$ 
indicates that the $^3\!P_2$ state is mainly populated by cascade feeding
which is also supported by a theoretical study \cite{Fritzsche2000}. 
The inset in Fig.~\ref{fig:ge_spectra} shows a zoom of the low-energy region of the spectrum.
The vertical lines in the inset mark the line energies and intensities obtained by a full cascade calculation
 \cite{Stohlker1999} whereas the continuous (red) line represents a fit of the theoretical model taking into account the experimental energy resolution.
 The good over-all correspondence between experiment and model gives confidence for the correct
identification of the strong line at the expected energy of the 
 $2\,^3\!P_2 \to 2\, ^3\!S_1$ transition. 

The measurement using the crystal spectrometer provided a much higher accuracy for the spectral line position. 
In this case, the observable energy range, 
 principally limited by the ion beam diameter and its distance 
from the crystal, was in the order of $4308±40$~eV -- much 
narrower than the linewidth observed with the Ge(i).
For an unambigous identification of the He-like uranium intra-shell transition,
the experimentally well known $1s^22p\, ^2\!P_{3/2} \to 1s^22s\, ^2\!S_{1/2}$ transition \cite{Beiersdorfer1993,Beiersdorfer1995} 
in Li-like uranium has been used as reference.
Similar to the He-like
system, Li-like ions were obtained by electron capture into He-like
uranium ions. 
In order to park the Li-like transition close to the He-like line, 
we used the Doppler effect selecting a 
kinetic energy of 32.63~MeV/u for Li-like ions (to be compared with 43.57~MeV/u for He-like ones).
This way both Li- and He-like uranium
transitions appeared at nearly the same Bragg angle 
and consequently in the same narrow spatial region of the position-sensitive detector.

The crystal spectrometer was mounted in the Johann geometry for the
detection of x rays with a corresponding Bragg angle
$\Theta_\textrm{B}$ around $46.0^\circ$. A cylindrically bent
germanium (220) crystal with size of $50 × 25$~mm$^2$ and a
radius of curvature of 800~mm was installed. The spectrometer did
not need any collimation because the imaging properties of the
curved crystal were used to resolve spectral lines from fast x-ray
sources nearly as well as for stationary sources \cite{Beyer1988}.
For this purpose it was necessary to place the Rowland-circle plane
of the spectrometer perpendicular to the ion-beam direction. In such
a configuration the spectral lines appear slanted in the image plane
of the spectrometer with their slope proportional to the ion-beam
velocity \cite{Beyer1988,Beyer1991,Trassinelli2007}.

Photons diffracted off the crystal were detected with a windowless
x-ray charge-coupled device (CCD), Andor DO420,  placed at 
$D$ = 575.75$±$0.65~mm 
away from the crystal, corresponding to a Bragg-angle setting of
$\Theta_\textrm{B}=46.0^\circ$. 
The spectrometer was kept under vacuum conditions (residual gas
pressure $\sim 10^{-5}$~mbar) to assure
proper working conditions for the CCD camera and to keep x-ray
absorption low. The interaction volume between the ion beam and the
nitrogen jet was located $758.05±0.46$~mm away from the crystal, \textit{i.e.} about
182~mm outside of the Rowland circle. The ion velocity was selected
to observe the x-ray radiation under study with a photon
energy of about 4308~eV in the laboratory frame, \textit{i.e.} in the
vicinity of the 8.6 keV K$\alpha_{1,2}$ lines of zinc observed in
the second order of diffraction.  At the same time, this matches the
spectrometer configured to a Bragg angle of $46^\circ$. The zinc
lines, produced by fluorescence with a commercial x-ray tube and a
zinc target, were used as calibration and for stability controls.
For this purpose a zinc plate was mounted on a removable support
between the target chamber and the crystal. In this configuration,
the spectrometer had an efficiency of $\sim7 ×
10^{-7}$ and a resolution of $\sim 2$~eV for the first order Bragg reflection. 

\begin{figure}[!tb]
\includegraphics[width=\columnwidth]{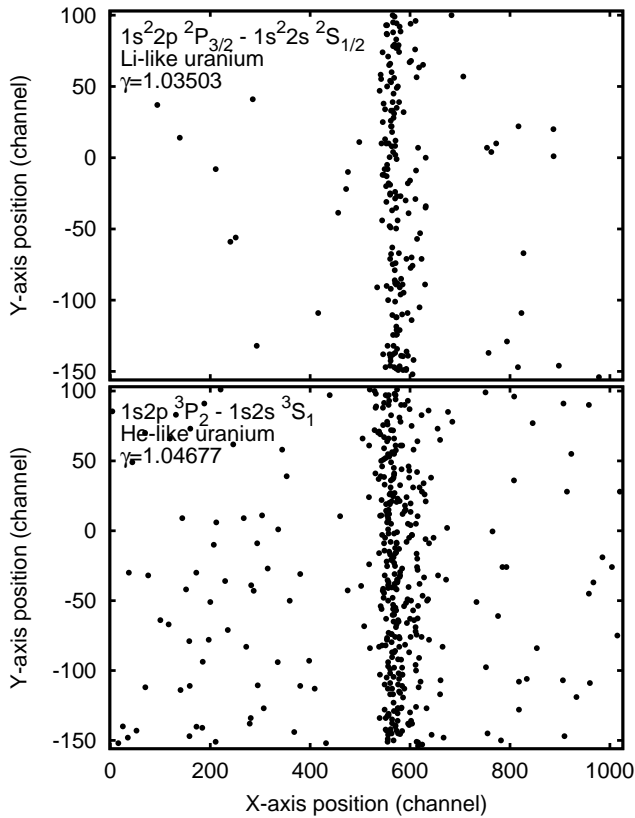}
\caption{Reflection of the Li-like  (up) 
and He-like (down) uranium intra-shell  transitions  on the Bragg spectrometer CCD. 
The transition energy increases with the increasing of x-position. 
The slightly negative slope of the line is due to the relativistic velocity of the fast ions.}
 \label{fig:2D}
\end{figure}

The data were acquired during a total period of about 4.5 days. 
To check for the stability of the spectrometer, daily calibrations with the zinc target 
were performed. 
During each accelerator cycle -- consisting of injection into the
ESR, cooling, deceleration, cooling, measurement -- the CCD acquired
data for 25 and 50 seconds for Li- and He-like U,
respectively. 
Over the whole experimental run, the spectrometer
was very stable, with a shift of $±3~\mu$m of the zinc-line
reflections on the CCD (corresponding to $±0.2$~pixels).
For the transition in He-like
uranium, a total number of about 300 counts in a net-time of 
$\sim 24$ hours was accumulated. For the Li-like ions, about 160 counts
in a net-time of $\sim 5$ hours were recorded.
As can be observed in Fig.~\ref{fig:2D}, the two intra-shell transitions could be 
identified unambiguously.
These spectra are characterized by a very low background constrained
only by the energy cuts and cluster analysis of the CCD raw data. 
No particular shields, other than the spectrometer stainless steel and
aluminum structure, were used.
We note that the shape of the lines corresponding to the fast ion emission is slightly asymmetric. Such
an asymmetry is, at present, not well understood. 
No relevant satellite transitions, due to a possible but improbable electron double capture \cite{Bednarz2003}, or parasitic fluorescence
lines could be found in this energy region, 
not even when higher reflection orders are considered.
Atomic cascade effects, like feeding and decaying time of the $1s2p\, ^3\!P_2$ level, which,  due to the Doppler effect,
could introduce an asymmetry in the x-ray energy distribution, are also unlikely to happen \cite{Fritzsche2000}.
How far a non-uniform x-ray reflectivity over the crystal surface 
\footnote{After the experiment, the crystal surface was surveyed by the x-ray optics group from the Institute for Optics and Quantum Electronics in Jena.} 
could cause irregularities in the spectral response of the spectrometer is difficult to judge. 

Using the Li-like intra-shell transition as
calibration, with energy  $E_\textrm{Li}= 4359.37±0.21$~eV
\cite{Beiersdorfer1993,Beiersdorfer1995}, the energy of the He-like
$2\,  ^3\!P_2 \to 2\, ^3\!S_1$ transition could be evaluated.
This reference line has been chosen instead of the K$\alpha_{1,2}$ zinc lines to reduce the
systematic uncertainty.
The accuracy of the observation angle, equal to $0.04^\circ$, is due to the spatial uncertainty of the fast-beam x-ray
source defined by the gas-jet and ion-beam positions. This caused a systematic uncertainty of about 
$±0.9$~eV on the fast-ion transition-energy measurement when the
Zn K$\alpha$ lines are used as reference. This problem can be
circumvented by using a calibration line originating from the fast
ion beam rather than from a stationary source.

Starting from Bragg's law in differential form, $\Delta E \approx - E
\cot \Theta_\textrm{B} \, \Delta \Theta_\textrm{B}$, one obtains an
approximate dispersion formula valid for small $\Delta
\Theta_\textrm{B}$. The energy of the He-like
$2\,  ^3\!P_2 \to 2\, ^3\!S_1$ transition  $E_\textrm{He}$ is given
by the simple formula
\begin{equation}
\frac{E_\textrm{He}}{ \gamma_\textrm{He}} \approx  \frac
{E_\textrm{Li}} {\gamma_\textrm{Li}} \left(  1 +  \frac {\Delta
x}{D\ \tan \Theta_\textrm{B}}   \right),
\end{equation}
where $\Delta x = x_\textrm{He} - x_\textrm{Li }$ denotes the
distance between the He- and Li-like U line images on the CCD and
$D$ the distance between crystal and CCD.
$\gamma_\textrm{He}=1.04677$ and $\gamma_\textrm{Li}=1.03503$ are
the Lorentz factors corresponding to the velocities of stored H- and
He-like ions, respectively. 
Their values are determined by the
accurately known voltages of the electron cooler.

\begin{figure}
\includegraphics[width=\columnwidth]{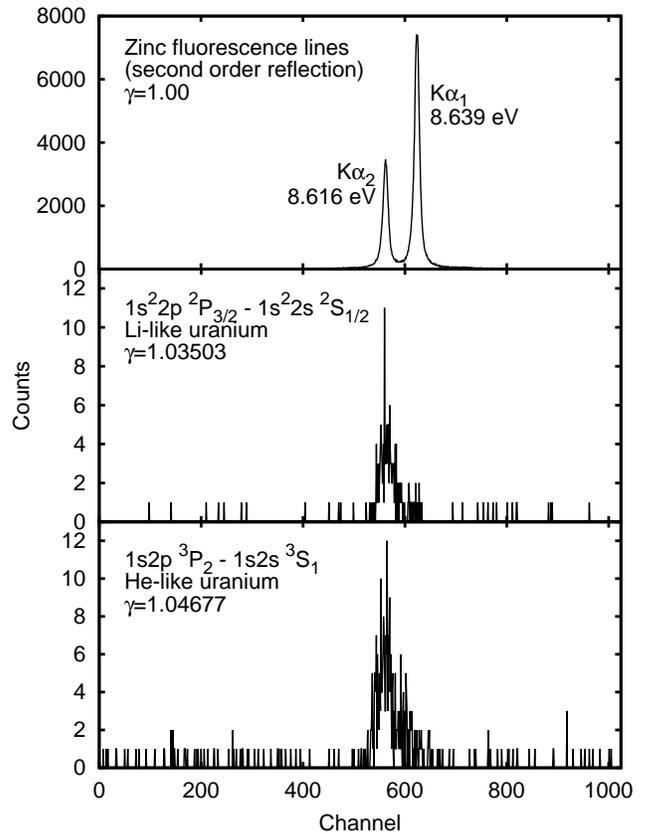}
\caption{X-ray spectra measured with the crystal spectrometer:
Zn K$\alpha$ doublet in second order (top)
and the intra-shell transitions of Li and He-like U (middle and bottom, respectively).}
 \label{fig:spectra}
\end{figure}

More complicated is the evaluation of the distance $\Delta x$ between the two
spectral lines from the fast ions, characterized by a slope due to the
Doppler effect \cite{Beyer1991}. Such a slope could not be
determined experimentally due to the low statistics and it was
calculated from simple geometrical considerations. The measured 
two-dimensional images have been rotated by the calculated slope
before projecting them on the spectrometer dispersion plane 
and measure the lines relative distance. 
The result of such  projections is presented in Fig~\ref{fig:spectra}.

Due to the low statistics,
no accurate investigation of the peaks asymmetry could be performed. 
The systematic effect induced by
this asymmetry has been estimated by comparing the line position
measurements obtained from different approaches. 
For this propose, a robust statistical analysis of the data set, like the median distribution, and a series of fit adjustment have been applied.
The median value was found in a self-consistent way regarding varying spectral intervals around the line center.
The fit adjustments were obtained via a likelihood function maximization for Poisson histograms \cite{Baker1984}, well adapted for low-statistics data sets, using different distributions such as Gauß-Cauchy curve, Gaussian convoluted with an exponential and a sum of two Gaussian. 
From this analysis, a scattering of $\approx 110~\mu$m of the $\Delta x$ value was measured, with a corresponding statistical uncertainty of about $65~\mu$m. 
The value of $\Delta x=+47~\mu$m, obtained via the median position calculation, has been considered as reference for further calculations.
The energy of the $1s2p\, ^3\!P_2 \to 1s2s\, ^3\!S_1$ transition was measured to be
\begin{equation}
E_\textrm{He} = 4509.71± 0.48_\text{stat} ± 0.83_\text{asym} ± 0.24_\text{syst}\; \text{eV},
\label{eq:result}
\end{equation} 
The first uncertainty is statistical, and the second one is due to
the peak asymmetry. The third term takes other sources of
systematic errors into account, of which the limited uncertainty
of the calibration line (0.21~eV) is the dominant one.
The uncertainty of the observation angle gives a contribution of $±$0.11~eV, drastically reduced compared to 
$±$0.9~eV when a stationary calibration source is used.
Other systematic uncertainties, including those originating from the beam velocity, are negligible (see ~\cite{Trassinelli2009}).

\begin{table}[!tb]
\caption{Comparison of the experimental transition energy, in eV,
for the $^3\!P_2 \to ^3\!S_1$ transition in He-like uranium with
various theoretical predictions.
The experimental uncertainty is obtained from the quadratic sum of the different uncertainties listed in Eq.~\eqref{eq:result}.}
\begin{center}
\begin{tabular}{l l}
Experiment & 4509.71 $±$ 0.99 \\
\hline
Theory \\
Indelicato 2008 \cite{Indelicato2008} & 4510.30 \\
Kozhedub 2008 \cite{Kozhedub2008} & 4509.86 $±$ 0.07 \\
Artemyev 2005 \cite{Artemyev2005} & 4510.03 $±$ 0.24 \\
Plante 1994 \cite{Plante1994} & 4510.46 \\
Chen 1993 \cite{Chen1993} & 4510.65 \\
Drake 1988 \cite{Drake1988} & 4510.01 \\
\hline
\end{tabular}
\end{center}
\label{tab1}
\end{table}
As presented in Table~\ref{tab1}, the measured 
transition energy for He-like uranium 
agrees well with  all theoretical predictions,
which, however, reflect different approaches.
The value provided by P.~Indelicato \cite{Indelicato2008}
represents a Multiconfiguration Dirac-Fock
(MCDF) calculation, obtained with the code developed by
P.~Indelicato and J.-P.~Desclaux \cite{Indelicato2005a}.
Radiative QED corrections $E^\text{QED}_\text{rad}$ to the inter-electronic interaction, \textit{i.e.}
QED screening effects, 
are taken into account via the self-consistent treatment 
of the one-loop vacuum polarization and the Welton approximation for the 
screened self-energy.
The value obtained by Artemyev et al. in \cite{Artemyev2005} is calculated {\it ab initio}. In \cite{Artemyev2005} all  QED corrections of first and second order in $\alpha$ are taken into account within the rigorous framework of QED. In addition to \cite{Indelicato2008} the non-radiative QED contribution to the inter-electronic interaction $E^\text{QED}_\text{non-rad}$, which is the difference between the contribution of the inter-electronic interaction diagrams calculated within the rigorous QED approach and within the Breit approximation, is calculated. Later in \cite{Kozhedub2008} the value from  \cite{Artemyev2005} has been improved using a new value of the mean-square nuclear-charge radius for uranium \cite{Kozhedub2008a} and the two-loop one-electron QED corrections for the excited states calculated in \cite{Yerokhin2006,Yerokhin2008}. Also in \cite{Kozhedub2008} the non-QED contribution of three- and more photon exchange diagrams 
has been calculated using the relativistic configuration-interaction method. 
Two-electron QED effects contribute to the $^3\!P_2$  $\to$  $^3\!S_1$  transition energy
with $E^\text{QED}_\text{2 el.}$ = $E^\text{QED}_\text{rad}$ + $E^\text{QED}_\text{non-rad}$ = 0.76~eV 
\cite{Artemyev2005}, which is close to the
present experimental uncertainty.

In addition to the absolute energy measurement of the $2\,  ^3\!P_2 \to 2\, ^3\!S_1$ transition,
the relative measurement 
of He- and Li-like uranium
intra-shell transitions could be evaluated.
In contrast with the previous case,
the experimental systematic uncertainty, peak asymmetry excluded,
reduced from 0.24 to 0.11~eV.
Theoretical uncertainties due to the finite nuclear size 
and the one-electron QED effects, are also drastically reduced. 
A comparison between our value  and different predictions is 
presented in Table~\ref{tab2}. 
As we can observe, all theoretical values agree 
with the experimental value within one standard deviation.

\begin{table}[!tb]
\caption{Experimental and theoretical energy difference between
He- and Li-like intra-shell transitions in eV. Contributions from one- and two-photon
exchange diagram, $E_\text{Breit}$, and from two-photon QED,
$E^\text{QED}_\text{rad}$ and $E^\text{QED}_\text{non-rad}$, are presented.
The experimental uncertainty is obtained from the quadratic sum of the different uncertainties.}
\begin{center}
\small
\begin{tabular}{l l l l l}
& $E_\text{Breit}$ & $E^\text{QED}_\text{rad}$ & $E^\text{QED}_\text{non-rad}$  & Total \\
\hline
Experiment &  &  & & 50.34 $±$ 0.96 \\
\hline
Theory \\
Indelicato \cite{Indelicato2008} & 51.61 & -1.66 & & 49.96 \\
Kozhedub \cite{Kozhedub2008} & 51.48 & -1.14 & -0.04 & 50.30 $±$ 0.03 \\
\hline
\end{tabular}
\end{center}
\label{tab2}
\end{table}

In summary, we report the first clear identification 
of the $1s2p\, ^3\!P_2 \to 1s2s\, ^3\!S_1$ transition
in He-like uranium. 
In addition we could measure the transition energy 
of such transition with a
relative uncertainty of $2×10^{-4}$, 
which is currently the most
accurate test of many-body and QED contributions in excited levels of 
very heavy He-like ions.
Differential measurements between different charge states of the
same fast ion  
pave the way for increased sensitivity via the reduction of the 
systematic uncertainty in both experimental and theoretical sides.
In the present experiment, the accuracy was  
principally limited by the statistics and the observed peak asymmetry of the fast-beam spectra. 
Hence, the improvement of the experimental set-up and an increazing of the 
acquisition time will allow 
in future experiments for more stringent tests
of the two-electron QED contributions in heavy highly charged ions.

\acknowledgments
We thank V.~Shabaev, A.N.~Artemyev  and A.~Surzhy­kov for
interesting discussions and theoretical support. 
We thank O. Wehrhan, H. Marschner and  E. Förster for the characterization of the spectrometer crystal.
The close
collaboration and support by the members of the ESR team, the A. von
Humboldt Foundation (M.T.), the DAAD (A.K., No.: A/05/52927) and I3
EURONS (EC contract no. 506065) are gratefully acknowledged.
This work was partially supported by Helmholtz Alliance HA216/EMMI.
Institut des Nanosciences de Paris and Laboratoire Kastler Brossel
are Unit{é} Mixte de Recherche du CNRS n$^{\circ}$ 7588 and
n$^{\circ}$ 8552, respectively.


\begin{thebibliography}{10}
\expandafter\ifx\csname url\endcsname\relax\def\url#1{\texttt{#1}}\fi

\bibitem{Persson1996}
\Name{Persson H., Salomonson S., Sunnergren P. \and Lindgren I.} \REVIEW{Phys.
  Rev. Lett. }{76}{1996}{204}.

\bibitem{Plante1994}
\Name{Plante D., Johnson W. \and Sapirstein J.} \REVIEW{Phys. Rev. A
  }{49}{1994}{3519}.

\bibitem{Artemyev2005}
\Name{Artemyev A.~N., Shabaev V.~M., Yerokhin V.~A., Plunien G. \and Soff G.}
  \REVIEW{Phys. Rev. A }{71}{2005}{062104}.

\bibitem{Marrs1995}
\Name{Marrs R.~E., Elliott S.~R. \and Stöhlker T.} \REVIEW{Phys. Rev. A
  }{52}{1995}{3577}.

\bibitem{Mokler2008}
\Name{Mokler P.~H., Lopez-Urrutia J. R.~C., Currell F.~J., Nakamura N., Ohtani
  S., Osborne C.~J., Tawara H., Ullrich J. \and Watanabe H.} \REVIEW{Phys. Rev.
  A }{77}{2008}{012506}.

\bibitem{Gumberidze2004}
\Name{Gumberidze A., Stöhlker T., Banas D., Beckert K., Beller P., Beyer H.~F.,
  Bosch F., Cai X., Hagmann S., Kozhuharov C., Liesen D., Nolden F., Ma X.,
  Mokler P.~H., Orsic-Muthig A., Steck M., Sierpowski D., Tashenov S., Warczak
  A. \and Zou Y.} \REVIEW{Phys. Rev. Lett. }{92}{2004}{203004}.

\bibitem{Schafer1989}
\Name{Schäfer A., Soff G., Indelicato P., Müller B. \and Greiner W.}
  \REVIEW{Phys. Rev. A }{40}{1989}{7362}.

\bibitem{Labzowsky2001}
\Name{Labzowsky L.~N., Nefiodov A.~V., Plunien G., Soff G., Marrus R. \and
  Liesen D.} \REVIEW{Phys. Rev. A }{63}{2001}{054105}.

\bibitem{Beiersdorfer1996}
\Name{Beiersdorfer P., Elliott S., Osterheld A., Stöhlker T., Autrey J., Brown
  G., Smith A. \and K. W.} \REVIEW{Phys. Rev. A }{53}{1996}{4000}.

\bibitem{Kukla1995}
\Name{Kukla K., Livingston A., Suleiman J., Berry H., Dunford R., Gemmell D.,
  Kanter E., Cheng S. \and Curtis L.} \REVIEW{Phys. Rev. A }{51}{1995}{1905}.

\bibitem{Martin1989}
\Name{Martin S., Buchet J.~P., Buchet-Poulizac M.~C., Denis A., Désesquelles
  J., Druetta M., Grandin J.~P., Hennecart D., Husson X. \and Lecler D.}
  \REVIEW{Eur. Phys. Lett. }{}{1989}{645}.

\bibitem{Indelicato1992}
\Name{Indelicato P., Birkett B.~B., Briand J.-P., Charles P., Dietrich D.~D.,
  Marrus R. \and Simionovici A.} \REVIEW{Phys. Rev. Lett. }{68}{1992}{1307}.

\bibitem{Toleikis2004}
\Name{Toleikis S., Manil B., Berdermann E., Beyer H.~F., Bosch F., Czanta M.,
  Dunford R.~W., Gumberidze A., Indelicato P., Kozhuharov C., Liesen D., Ma X.,
  Marrus R., Mokler P.~H., Schneider D., Simionovici A., Stachura Z., Stöhlker
  T., Warczak A. \and Zou Y.} \REVIEW{Phys. Rev. A }{69}{2004}{022507}.

\bibitem{Munger1986}
\Name{Munger C.~T. \and Gould H.} \REVIEW{Phys. Rev. Lett. }{57}{1986}{2927}.

\bibitem{Fritzsche2000}
\Name{Fritzsche S., Stöhlker T., Brinzanescu O. \and Fricke B.}
  \REVIEW{Hyperfine Interactions }{127}{2000}{257}.

\bibitem{Ma2001}
\Name{Ma X., St\"ohlker T., Bosch F., Brinzanescu O., Fritzsche S., Kozhuharov
  C., Ludziejewski T., Mokler P.~H., Stachura Z. \and Warczak A.} \REVIEW{Phys.
  Rev. A }{64}{2001}{012704}.

\bibitem{Eichler2007}
\Name{Eichler J. \and Stohlker T.} \REVIEW{Phys. Rep. }{439}{2007}{1}.

\bibitem{Indelicato2005a}
\Name{Indelicato P. \and Desclaux J.} \Book{{MCDFGME}, a {MultiConfiguration
  Dirac Fock} and general matrix elements program (release 2005)}
  \url{http://dirac.spectro.jussieu.fr/mcdf} (2005).

\bibitem{Stohlker1999}
\Name{Stöhlker T.} \REVIEW{Phys. Scripta T}{80A}{1999}{165}.

\bibitem{Beiersdorfer1993}
\Name{Beiersdorfer P., Knapp D., Marrs R.~E., Elliott S.~R. \and Chen M.~H.}
  \REVIEW{Phys. Rev. Lett. }{71}{1993}{3939}.

\bibitem{Beiersdorfer1995}
\Name{Beiersdorfer P.} \REVIEW{Nucl. Instrum. Meth. Phys. Res. B
  }{99}{1995}{114}.

\bibitem{Beyer1988}
\Name{Beyer H.~F. \and Liesen D.} \REVIEW{Nucl. Instrum. Meth. Phys. Res. A
  }{272}{1988}{895}.

\bibitem{Beyer1991}
\Name{Beyer H.~F., Indelicato P., Finlayson K.~D., Liesen D. \and Deslattes
  R.~D.} \REVIEW{Phys. Rev. A }{43}{1991}{223}.

\bibitem{Trassinelli2007}
\Name{Trassinelli M., Banas D., Beyer H.~F., Jagodzinski P., Kumar A., Pajek M.
  \and St\"ohlker T.} \REVIEW{Can. J. Phys. }{85}{2007}{441}.

\bibitem{Bednarz2003}
\Name{Bednarz G., Sierpowski D., Stohlker T., Warczak A., Beyer H., Bosch F.,
  Brauning-Demian A., Brauning H., Cai X., Gumberidze A., Hagmann S.,
  Kozhuharov C., Liesen D., Ma X., Mokler P.~H., Muthig A., Stachura Z. \and
  Toleikis S.} \REVIEW{Nucl. Instrum. Meth. Phys. Res. B }{205}{2003}{573}.

\bibitem{Baker1984}
\Name{Baker S. \and Cousins R.~D.} \REVIEW{Nucl. Instrum. Meth. Phys. Res.
  }{221}{1984}{437}.

\bibitem{Trassinelli2009}
\Name{Trassinelli M., Kumar A., Beyer H.~F., Indelicato P., Martin R., Reuschl
  R. \and Stohlker T.~H.} \REVIEW{J. Phys. CS }{163}{2009}{012026}.

\bibitem{Indelicato2008}
\Name{Indelicato P.} unpublished (2008).

\bibitem{Kozhedub2008}
\Name{Kozhedub Y. \and Shabaev V.} unpublished (2008).

\bibitem{Chen1993}
\Name{Chen M.~H., Cheng K.~T. \and Johnson W.~R.} \REVIEW{Phys. Rev. A
  }{47}{1993}{3692}.

\bibitem{Drake1988}
\Name{Drake G.} \REVIEW{Can. J. Phys. }{66}{1988}{586}.

\bibitem{Kozhedub2008a}
\Name{Kozhedub Y.~S., Andreev O.~V., Shabaev V.~M., Tupitsyn I.~I., Brandau C.,
  Kozhuharov C., Plunien G. \and Stohlker T.} \REVIEW{Phys. Rev. A
  }{77}{2008}{032501}.

\bibitem{Yerokhin2006}
\Name{Yerokhin V.~A., Indelicato P. \and Shabaev V.~M.} \REVIEW{Phys. Rev.
  Lett. }{97}{2006}{253004}.

\bibitem{Yerokhin2008}
\Name{Yerokhin V.~A., Indelicato P. \and Shabaev V.~M.} \REVIEW{Phys. Rev. A
  }{77}{2008}{062510}.

\end{thebibliography}

\end{document}